# Fundamental Diagram of Pedestrian Dynamics by Safety Interspace Model[*]

Jun Fang (方峻)[a)†], Zheng Qin (覃征)[a)], Zhengcai Lu (卢正才)[a)], Fengfei Zhao (赵凤飞)[a)]

a) Department of Computer Science and Technology, Tsinghua University, Beijing 100084, China

**Abstract:**

The NaSch model is a classical one-dimensional cellular automata (CA) model of freeway traffic. However, it is not suitable for simulating pedestrian dynamics due to the kinetic difference between vehicles and pedestrians. Based on NaSch model, we proposed a new safety interspace model to investigate the single-file pedestrian stream. We simulated and reproduced Seyfried's experiment in 2006 quantitatively and use the empirical data to validate our model. We defined the safety interspace as a linear function of current velocity and introduced a normal distributed random variable to describe the randomness of spatial cognition. Meanwhile, we extended NaSch model in space partition from single-grid mode to multi-gird mode to increase spatial resolution. It was found that through three model perimeters with appropriate values, the new model could reproduce faithfully the typical form of the fundamental diagram from empirical data. The velocity-density curve in simulation could indicate the discontinuity areas in fundamental diagram observed from empirical data. In addition, we reproduced the stop-and-go waves and phase separation into a stopping area and an area where pedestrians walk slowly at high density.

---

[*] Project supported by Specialized Research Fund for the Doctoral Program of Higher Education of China (Grant No 20090002110085).

[†] Corresponding author. Email: fangjun06@mails.tsinghua.edu.cn

# 1 Introduction

In recent years, the modeling of pedestrian flows has become one of the most exciting topics. It has attracted considerable attention from the physical science, traffic engineering, computer engineering, or even social psychology science.[1] Besides the continuous space models (e.g. the social force model[2] and the adaptive velocity model[3, 4]), the discrete space models have also been proposed and researched extensively, especially the cellular automata (CA) model.[5] The CA model is a simple but effective method to simulate the movement of a large number of pedestrians. At very low cost in simulation time, it is able to simulate the empirical results qualitatively and quantitatively with acceptable error. The CA model can reproduce several typical self-organization phenomena in controlled experiments and pedestrian traffic, such as the phase transition, phase separation, faster-is-slower effect, stop-and-go wave and lane formation.[1, 6]

Although a large variety of models for pedestrian dynamics has been proposed, there have been only limited attempts to calibrate and validate these approaches so far. Since 2005, Armin Seyfried and his colleges in Jülich Supercomputing Centre hired many participants and carried out a series of well-controlled laboratory experiments.[7] They collected data via automated determination of trajectories,[8][9] analyzed data [10] and developed models to describe the dynamic of pedestrians quantitatively.[3, 4, 11, 12] Their work is a part of the Herms project [13] in which the data resulting from the experiments will be used to calibrate and test pedestrian movement models.

Among several experimental scenarios investigated by Seyfried's group, the scenario of single-file movement is a basic setup and has been carefully researched. Although this scenario is rare in real life, it is suitable for experimental research. This experiment setup can reduce uncontrollable influences as much as possible and make it possible to study the influence of single parameters, like the width, one-dimensional density and velocity. So far, Seyfried's group modeled the single-file movement only by continues space models.[3, 4, 14] In this study, we try to investigate the single-file movement by a new discrete space model, which is called safety interspace model. We describe the dynamic properties of single-file pedestrian streams quantitatively and use empirical data from Seyfried's experiment [7] to validate our model.

Our safety interspace model is based on the NaSch (NS) model.[15] The NaSch model is a type of classical one-dimensional CA model, which was used to model freeway traffic of one-lane.[16, 17] There are some similar self-organization phenomena between vehicle flow and pedestrian flow, such as the transition from laminar traffic flow to stop-and-go waves with increasing density. However, the differences between vehicle and pedestrian in kinetic character are obvious. The pedestrian movements are more flexible and chaotic than vehicle. Pedestrians are capable of changing speed more quickly when gaps arise. They can accelerate to full speed from a standstill and halt from full speed immediately. We made some considerable modifications to NaSch model to adapt the new model for pedestrian simulation. In addition, we extended NaSch model in space partition from single-grid mode to multi-gird or finer discretization mode to increase spatial resolution. The multi-grid CA model is often adopted in pedestrian simulation.[18, 19] It can describe pedestrian and walk environment more accurately, and increase the computational accuracy of position and velocity.

## 2 Models

### 2.1 Standard NaSch Model

The NaSch model is a type of classical one-dimensional CA model, which was used to simulate highway traffic of single lane. One update of the system consists of the following three consecutive steps, which are performed in parallel for all vehicles:[15]

(1) Acceleration:

$$v_i^{n+\frac{1}{2}} = \min\{v_i^n + 1, V_{\max}, d_i^n\} \qquad (1)$$

Where $v_i^n$ is the velocity of vehicle $i$ at time step $n$, which is an integer between 0

and $V_{max}$. $d_i^n$ denotes the interspace between vehicle $i$ and its front vehicle $i-1$. $d_i^n = x_{i-1}^n - x_i^n - L_p$ where $L_p$ is the length of vehicle.

(2) Random slowing:

$$v_i^{n+1} = (1-p)v_i^{n+\frac{1}{2}} + p \cdot \max\{v_i^{n+\frac{1}{2}} - 1, 0\} = \max\{v_i^{n+\frac{1}{2}} - p, 0\} \quad (2)$$

Where $p$ denote the probability of random slowing. It takes into account natural velocity fluctuations caused by human behavior and/or by varying external conditions.

(3) Motion:

$$x_i^{n+1} = x_i^n + v_i^{n+1} \quad (3)$$

The vehicle $i$ is advanced $v_i^n$ sites and update its position at time step $n+1$.

## 2.2 Modifying NaSch model for pedestrian dynamics

The differences between vehicle and pedestrian in kinetic character are obvious. The velocity of pedestrian is no more than 1.4m/s under easy mind.[20] Pedestrians are capable of changing speed more quickly when gaps arise. They can accelerate to full speed from a standstill as well as halt from full speed immediately. Therefore, the rule of velocity acceleration in NaSch model (eq. (1)) is not suitable for pedestrian modeling. In our model, we suppose the pedestrian velocity can be changed with any amplitude ($\leq V_{free}$) during one time step.

The changes of human physiological and psychological states under easy mind will cause natural fluctuation of walking speed, even in the range of low density. The fluctuation of walking speed is somewhat similar to the random slowing of vehicle. However, the rule of random slowing in NaSch model is difficult to be defined and quantified in our model. To our knowledge, the random slowing in pedestrian dynamics has not reported and discussed in published documents. Due to data missing, we cannot determine the value of slowing probability and decreasing amplitude of velocity. For simplicity, we replace decreasing amplitude of velocity with a random variable $g_i^n$.

Velocity update in standard NaSch model is replaced by the following two equations:

$$g_i^n = \max\{N(0, \sigma^2), 0\} \tag{4}$$

$$v_i^{n+1} = \min\{\max[d_i^n - g_i^n, 0], V_{free}\} \tag{5}$$

Where $g_i^n$ is normal distributed with mean 0 and variance $\sigma^2$. A person does not move to the closest empty cell but leave a distance $g_i^n$ as his safety interspace to its front person. Following this idea, we proposed a more general model called the safety interspace model in the next subsection. The above modified model is a special case of the safety interspace model with $k = 0, \mu = 0$.

## 2.3 Safety interspace model

We extended above modified model and proposed a new model called the safety interspace model. In pedestrian walking, the back person will keep a distance to front person for safety and comfort. Unlike vehicle traffic, the hit between persons is not life-threatening but painful or uncomfortable. The requirement of safety interspace for person is flexible and fuzzy. The safety interspace of each one is sensitive to his own velocity and varying depending on his physiological and psychological states. The safety interspace in our model is calculated by a linear function $g = k \cdot v + \xi$, which can be divided into two items. The first item $g_1 = k \cdot v$ is directly proportional to current velocity. The velocity is larger; the safety interspace reserved is longer. It is similar to the time-headway maintaining in vehicle traffic. The second item $g_2 = \xi$ is a random variable to display the perceptive randomness of human. For the difference of physiological and psychological states, the spatial cognition is fluctuating during a certain range for different individuals and/or different time. $\xi$ is designed as a normal distributed random variable with mean $\mu$ and variance $\sigma^2$, where $\mu$ denotes the basic safety distance at a standstill and $\sigma$ control the amplitude of fluctuation. As an experimental evidence, the linear function $\beta = 0.125 + 0.756v$ gives the best fit to the empirical data in [4, 21] about the relation between the safety distance and velocity. For the safety interspace model, Eq. (4) is changed to be:

$$g_i^n = \max\{k \cdot v_i^n + N(\mu, \sigma^2), 0\} \tag{6}$$

There are three model parameters: slope $k$, mean $\mu$ and standard deviation $\sigma$. We will validate our model using empirical data [21] and test whether any of the parameters can be dispensable. We will show that through the three model perimeters with appropriate values, our new model can reproduce faithfully the typical form of the fundamental diagram from empirical data.

## 3 Simulation and Discussion

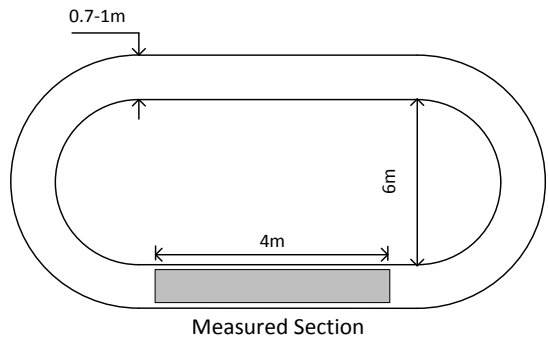

Fig. 1 Sketch of the experimental setup arranged in 2006 in the wardroom of the Bergische Kaserne Düsseldorf.[7]

Our model is proposed to simulate and reproduce the Seyfried's experiment quantitatively, which is performed in 2006 in the wardroom of the Bergische Kaserne Düsseldorf with a test group of up to 70 soldiers.[7, 11, 21] We validate our model by the empirical data of that experiment. The length of the circular passage was about 26m, with a $l = 4$m measured section (see Fig. 1). The testers are instructed to keep single-file movement and overpass is not allowed.

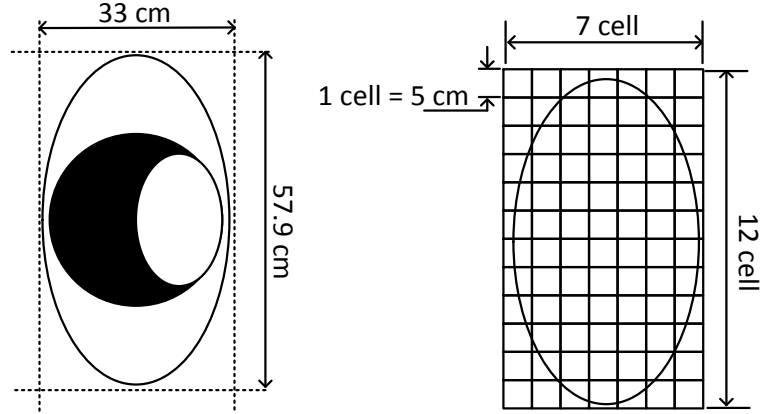

Fig. 2 The standard projected area of pedestrian in [22] and its spatial discretization in our model.

We used a virtual passage with periodic boundary and a virtual measurement equipment to simulate the experimental setup in.[7] We use the multi-grid CA model to discretize simulation environment. According to [22], the standard projected area of an adult is 0.33m*0.579m. We discretize the projected area using 7cell*12cell whose unit length is 0.05m, i.e. the size of virtual person is 0.35m*0.6m (see Fig. 2). According to this granularity, the length of passageway is 520cell (see Fig. 3) and can hold 74 persons at most, i.e. maximum one-dimensional density $\rho_{1d} = 2.86$ped./m. The unit time step in our simulation is 0.5s and the resolution of velocity is 0.1m/s accordingly. We set the free velocity $V_{free} = 1.3$m/s in simulation. Only the velocity of horizontal movement is considered, regardless of vertical movement.

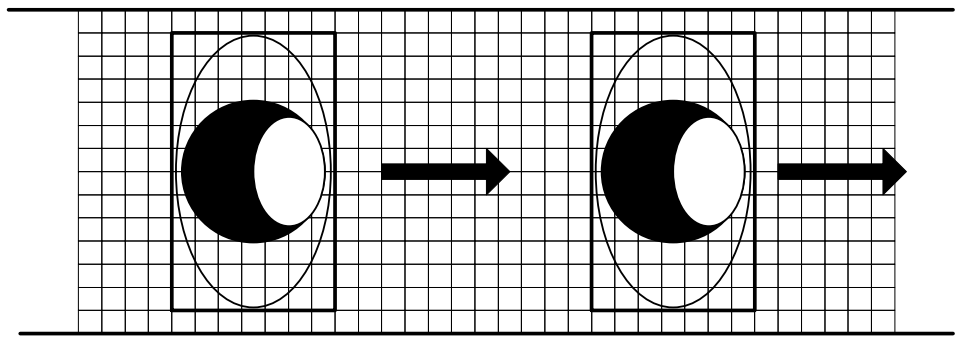

Fig. 3 Single-file movement in the discrete space of passageway.

Except for some special number, such as $N = 10$ and $N = 20$, persons in simulation cannot be uniformly distributed in the discretized passageway. At initial stage, we put all virtual persons one by one in the passageway with no gap between them. We test different number of testers from $N = 2$ to $N = 70$ with increasing by one person each simulation. Each simulation run 10,000 time steps and the 5,001~10,000 steps is considered as the steady state.

Similar to the statistic method of [21], the individual velocity $v_i$ of person $i$ is calculated by

$$v_i = \frac{l_m}{n_i^{out} - n_i^{in}} \quad (7)$$

where $n_i^{in}$ and $n_i^{out}$ is the entrance and exit times of measured section for pedestrian $i$ and $l_m$ is the length of measured section. The instantaneous density in the measured section is

$$\rho(n) = \sum_{i=1}^{N} \frac{\Theta_i(n)}{l_m} \quad (8)$$

Where $N$ is the total number of person and $\Theta_i(n)$ measures the fraction of space between person $i$ and $i+1$ inside the measurement area:

$$\Theta_i(n) = \begin{cases} \dfrac{n - n_i^{in}}{n_{i+1}^{in} - n_i^{in}} : n \in [n_i^{in}, n_{i+1}^{in}] \\ 1 : n \in [n_{i+1}^{in}, n_i^{out}] \\ \dfrac{n_{i+1}^{out} - n}{n_{i+1}^{out} - n_i^{out}} : n \in [n_i^{out}, n_{i+1}^{out}] \\ 0 : \text{otherwise} \end{cases} \quad (9)$$

The associated individual density $\rho_i$ is the average of $\rho(n)$ during $n_i^{in}$ and $n_i^{out}$:

$$\rho_i = \frac{\sum_{n_i^{in}}^{n_i^{out}} \rho(n)}{n_i^{out} - n_i^{in}} \quad (10)$$

We firstly investigate the fundamental diagram from the relationship between velocity and density in global view. For each value of density, we calculate the ensemble average velocity by averaging individual velocity of all persons. We will investigate every model parameter ($k, \mu$ and $\sigma$ in Eq. (6)) independently to discuss its control over pedestrian dynamics.

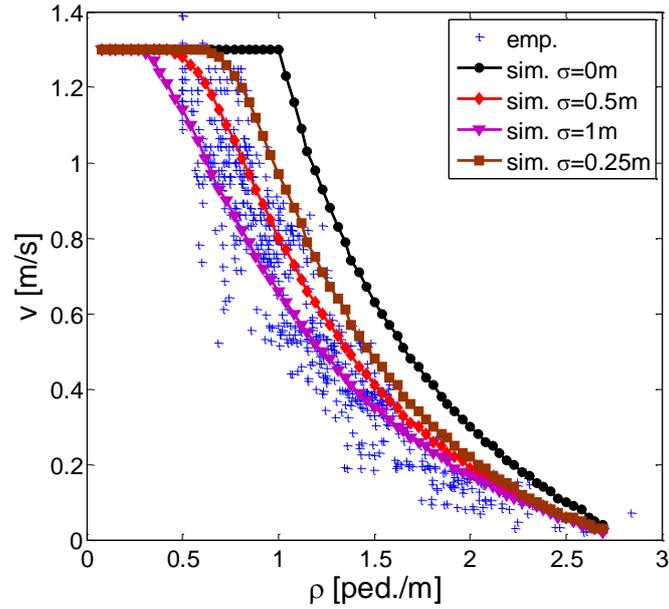

Fig. 4 Velocity-density relation of safety interspace model with $k = 0$, $\mu = 0$ in comparison with the empirical data.

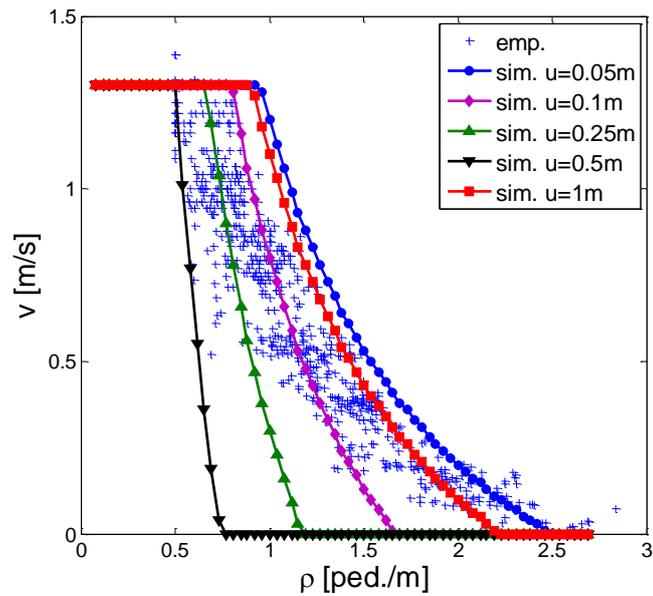

Fig. 5 Velocity-density relation of safety interspace model with $k = 0$, $\sigma = 0$ in comparison with the empirical data.

First, we investigate the standard deviation $\sigma$ of safety interspace model in Eq. (6) (see Fig. 4). It is found that in low density $\rho < 1.12$ ped./m, the velocity-density curve with $\sigma = 0.5$m fit the empirical data well. However, in high density $\rho > 2$ ped./m, the average velocity of four case with different σ values is higher than empirical data. In the density $\rho = 2.69$ ped./m ($N = 70$), all curves converge at a point. With increasing σ, the critical density becomes smaller and the slope of curve decreases.

Therefore, $\sigma$ acts on the average velocity mainly in the range of low density $\rho < 1.12$ ped./m.

Unlike $\sigma$, $\mu$ controls velocity in the total range of density, especially at high density (see Fig. 5). For $\sigma = 0$, the model is deterministic. Without fluctuation of safety interspace, the movable requirement $d > \mu$ is rigid to make average velocity fall down to zero quickly with increasing density. With increasing density, the critical density becomes smaller and the slope becomes larger. The result of $\mu = 0.1m$ fit the empirical data best among the five curves from the eye observation. This result coincides with [21] that the safety distance $\beta = 0.125 + 0.758\,v$. For $k = 0$ and $\sigma = 0$ in Eq. (6), we cannot realize $\mu = 0.125$m but $\mu = 0.1$m in our simulation as the one of the closest values. ‡

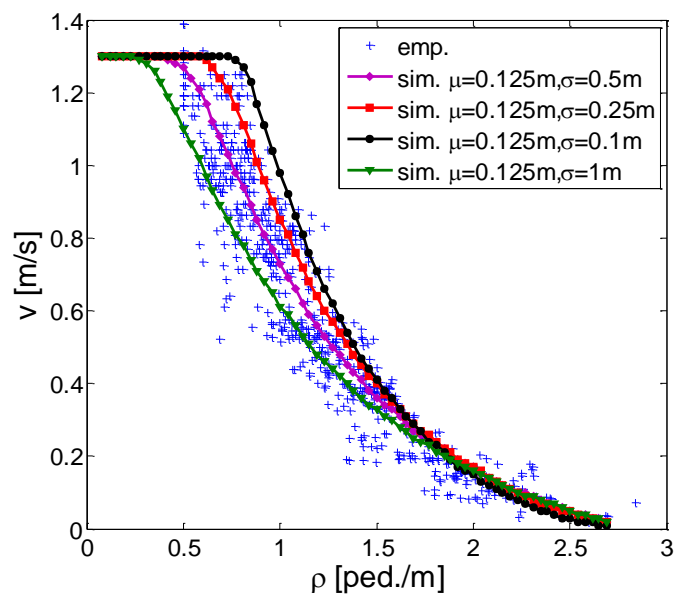

Fig. 6 Velocity-density relation of safety interspace model with $k = 0$ in comparison with the empirical data.

In Fig. 6, we adopt $\mu = 0.125m$ according to the function of safety distance $\beta = 0.125 + 0.758\,v$ in [21] and compare different $\sigma$ value. We find the curve with $\sigma = 0.5m$ fits the empirical data best, coinciding with Fig. 4. It seems that safety interspace model with only two parameters $\mu$ and $\sigma$ can be enough to reproduce the typical form of the fundamental diagram. However, $k$ is indispensable for catching the nature of pedestrian dynamics, which will be analyzed in the following.

---

‡ Due to discrete space, each person can only move an integral number of cell. The real valued $g_i^n$ must be rounded to an integral value. In simulation program, we adopt the banker's rounding. For example, 1 cell = 0.05m and 2.5 cell = 0.125m. Therefore, rounding $\mu = 2.5$ cell becomes $\mu = 2$ cell, i.e. 0.1m.

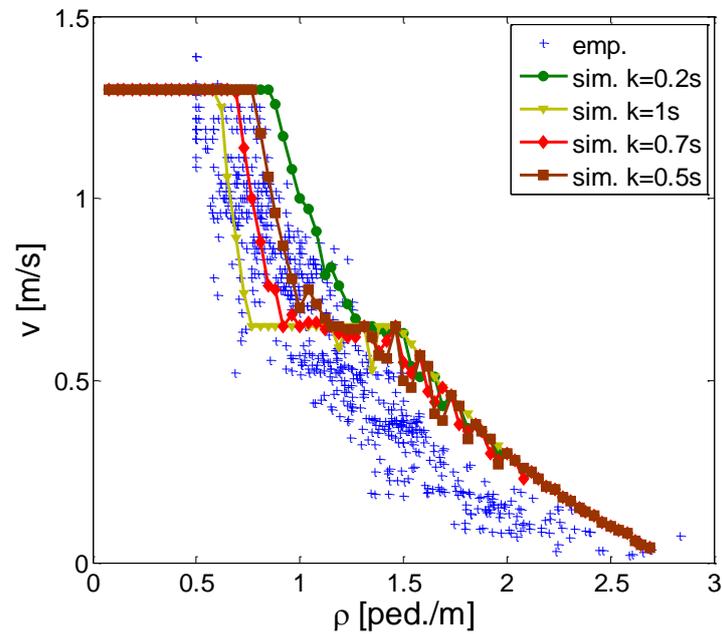

Fig. 7 Velocity-density relation of safety interspace model with $\mu = 0$ and $\sigma = 0$ in comparison with the empirical data.

Next, we investigate the slope $k$ of velocity in safety interspace equation (see Fig 7). We observe the curve of $k = 0.7$ lies the middle of empirical data for $\rho < 0.92$ ped./m, which coincides with $\beta = 0.125 + 0.758\, v$ in [21]. However, it gets close gradually and finally merges with other curves after $\rho = 0.92$ ped./m. There is a clear dividing line at $v = 0.65$ m/s. Above the line, each curves of different $k$ falls down and keeps apart with increasing density. Below the line, all curves get close and merge together finally. $k$ controls average velocity effectively in the range of low density but loses its control in the range of high density.

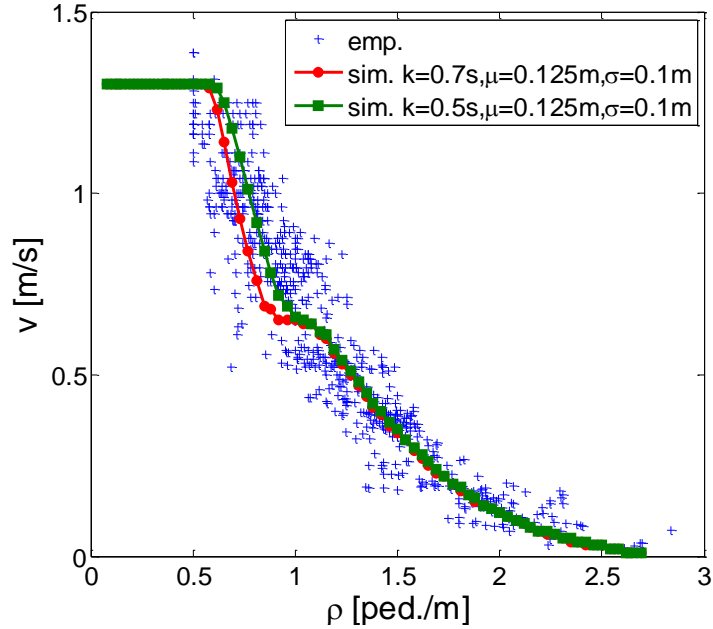

Fig. 8 Velocity-density relation of safety interspace model with different $k$ and same $\mu$ and $\sigma$ in comparison with the empirical data.

We have test different value sets of parameter $k, \mu,$ and $\sigma$, and found the value set of $k = 0.5\text{s}, \mu = 0.125\text{m}, \rho = 0.1\text{m}$ fit the empirical data best, which is even better than $k = 0.7\text{s}, \mu = 0.125\text{m}, \rho = 0.1\text{m}$. This simulation result about the value of $k$ has a little deviation from the conclusion $\beta = 0.125 + 0.758\,v$ in [21] (see Fig. 8).

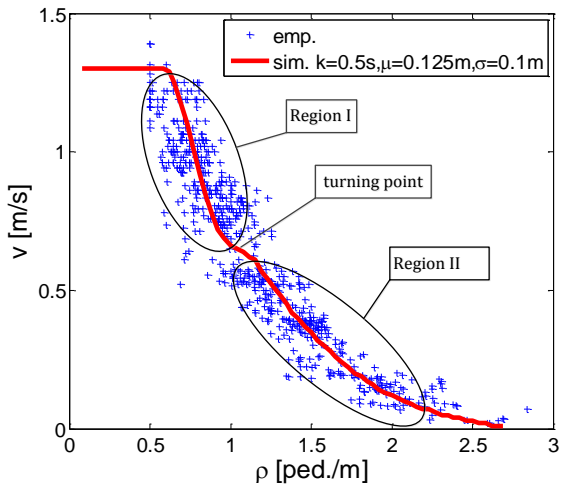
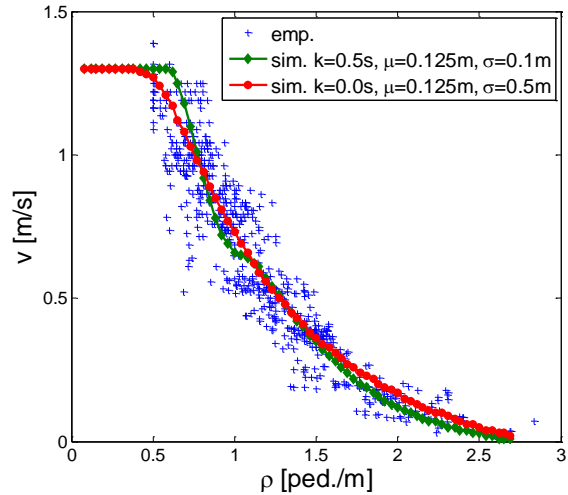

Fig. 9 (a) Discontinuity of the fundamental diagram.

Fig. 9 (b) Comparison of two simulation results with and without $k$.

The model parameter $k$ is indispensable for catching the nature of pedestrian dynamics. From Fig. 9(a), there is a discontinuity of the fundamental diagram when the density is around 1ped./m and velocity is around 0.65m/s ($0.5\text{V}_{\text{free}}$). This phenomenon has been observed in field experiment in [23] and can be explained by the boundary –induced phase transition in [24][25]. The discontinuity in empirical data can be reproduced by our simulation model with appreciate values of parameters. In the middle of velocity-density curve, there is an turning point in $\rho = 1$ped./m. It separates the fundamental diagram into two areas. One is the area of $\rho < 1$ped./m with negative curvature and another is the area of $\rho > 1$ped./m with positive curvature (see Fig. 9(a)). From Fig. 9(b), the two curves with and without $k$ seem to be similar in reproducing $v - \rho$ fundamental diagram. However, the model without $k$ cannot reproduce the discontinuity. Its $v - \rho$ curve has not the turning point and its curvature is negative in the total range of density. It falls more slowly than the curve with $k$ in the region-I. Therefore, the three parameters are indispensable in our model to reproduce empirical data.

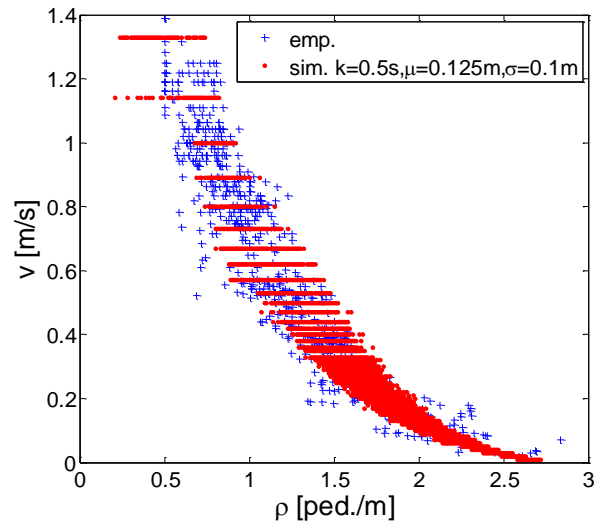

Fig. 10 Individual velocity-density data from virtual measured section in comparison with empirical data.

From virtual measured section in simulation, we obtain individual velocity-density data and plot a scatter diagram in comparison with empirical data (see Fig. 10). One find the simulation results can override the major area of empirical data. However, the velocity values in simulation do not distribute uniformly and many values of empirical data are not reproduced in the range of low density $\rho < 1$ ped./m. This can be explained by the following two reasons. One is that the values of position and velocity are both integer due to discrete space of model. The real valued safety interspace $g_i^n$ in Eq. (6) must be rounded in the step of motion, which will generate the round-off error. The round-off error of position is ±0.05m and that of velocity is ±0.1m/s. Another reason is that our simulation adopt the same parameters $k, \mu$ and $\sigma$ for all persons while the adaptive velocity model [3, 4, 21] adopt normal distributed individual parameters $a_i, b_i, \tau_i$ for person $i$. We only use a normal distribution item $\xi = N(\mu, \sigma^2)$ in safety interspace to describe the perceptive randomness. The standard deviation $\sigma$ controls the amplitude of fluctuation and takes quite limited effect in the range of low density for enough space between two successive cars.

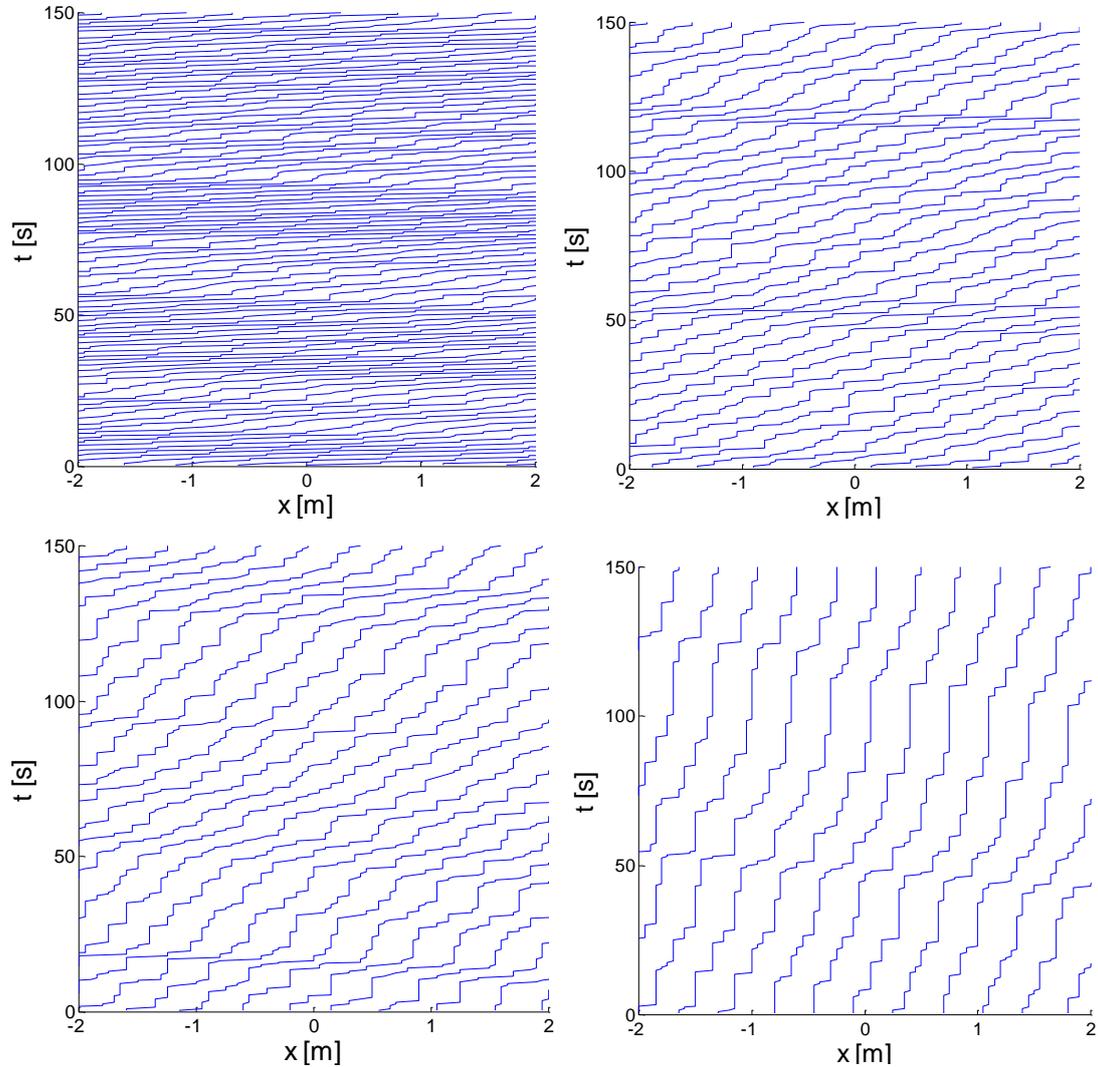

Fig. 11 Trajectories for the runs with $N = 39, 56, 62, 70$ (left to right, top to bottom) in simulation with $k = 0.5s, \mu = 0.125m, \sigma = 0.1m$. The direction of movement is right.

In Fig. 11, the x-component of trajectories is plotted against time obtained from the virtual measured section of $l_m = 4$m in simulation. With increasing density, the stop-and-go waves accumulate gradually and propagate opposite to the movement direction (upstream). Phase separation appears in the modeled trajectories as in field experiments.[21] Macroscopically one observes the phase separation into a stopping area and an area where pedestrians walk slowly. However, pauses lasting one time step frequently occur in our simulation for $N = 39$ while stopping is first observed during the runs with $N = 45$ in empirical data in.[21] This is because our model does not consider the velocity effect or anticipation effect in parallel update. In this effect, the back person calculates the interspace considering only the current position of front person, whether the front person move ahead or not next time. This will be investigated in more detail in a separate publication.

## 4 Conclusion

In this study, we propose a CA model called the safety interspace model to describe the pedestrian's single-file movement. Our model is based on the NaSch model and we made some considerable modification to adapt our new model for the pedestrian simulation. We simulate and reproduce Seyfried's experiment in [21] quantitatively and use its empirical data to validate our model. In our model, a person does not move to the closest empty cell but leave a distance as the safety interspace to its front person. The safety interspace in our model is calculated by a linear function of current velocity. In this function, we introduce a normal distributed random variable to describe the perceptive randomness of safety distance. In addition, we extended NaSch model in space partition from single-grid mode to multi-gird mode, in order to increase the spatial resolution of discrete model.

The results of simulation experiments includes: (a) the relationship of global average velocity to global density in fundamental diagram; (b) individual velocity-density data from measured section; (c) shock wave propagation and phase separation. The simulation results have shown that through three model perimeters with appropriate values, the new model can reproduce faithfully the typical form of the fundamental diagram from empirical data. The three parameters are indispensable. Without one of them, the model cannot reproduce the discontinuity phenomena in fundamental diagram observed in field experiment.[23] In addition, we reproduce the propagation of stop-and-go waves and phase separation into a stopping area and an area where pedestrians walk slowly at high density.

There are some imperfections for our current model. First, our simulation adopt the same parameters $k, \mu$ and $\sigma$ for all persons while the adaptive velocity model [3, 4, 21] adopted normal distributed individual parameters $a_i, b_i$ and $\tau_i$ for person $i$. We only use a normal distribution item $\xi = N(\mu, \sigma^2)$ to describe the perceptive randomness of safety interspace. The model parameter $\sigma$ controls the amplitude of fluctuation and takes quite limited effect in the range of low density for enough space between two successive cars. Second, we do not consider the anticipant effect in parallel update. In this effect, the back person calculates the interspace considering only the current position of front person, whether the front person move ahead or not next time. This will be investigated in more detail in a separate publication.

## Acknowledgements

We are grateful to Armin Seyfried in Jülich supercomputing centre and Jun Zhang in Bergische Universität Wuppertal for providing very useful empirical data and suggestion. We thank department of computer simulation for fire safety and pedestrian traffic in bergische universität Wuppertal for their web pages with experimental data and video. We use the data that come from the project funded by the German Science Foundation (DFG) under DFG-Grant No. KL 1873/1-1 and SE 1789/1-1.